\documentclass[floats,aps,twocolumn,showpacs]{revtex4}
\usepackage{dcolumn}
\usepackage{epsfig}
\newcommand {\la} {\langle}
\newcommand {\ra} {\rangle}
\newcommand {\beq} {\begin{eqnarray}}
\newcommand {\eeqn} [1] {\label{#1} \end{eqnarray}}%

\newcommand {\ve} [1] {\mbox{\boldmath $#1$}}

\begin{document}
%
%\initfloatingfigs
%

%\tighten
\title{
A relation between proton and neutron asymptotic normalization coefficients for  
light mirror nuclei and its relevance to nuclear astrophysics.}

\author{
N.\ K.\ Timofeyuk$^{1)}$, R.C. Johnson$^{1)}$ and A.M. Mukhamedzhanov$^{2)}$
}                 

\affiliation{
$^{1)}$ Department of Physics, %School of Electronics and Physical Sciences, 
University of Surrey, Guildford,
Surrey GU2 7XH, England, UK\\
$^{2)}$Cyclotron Institute, Texas A$\&$M University, College Station, TX 77843
}

%\author{
%A.M. Mukhamedzhanov
%\address{
%Cyclotron Institute, Texas A$\&$M University, College Station, TX 77843
%}

\date{\today}

\begin{abstract}

We show how  the charge symmetry of strong interactions can be used to 
relate the proton and neutron asymptotic normalization coefficients (ANCs) 
 of the one nucleon overlap integrals for light mirror nuclei.
This relation extends to the case of real proton decay where
 the mirror
analog  is a virtual neutron decay of a loosely bound state.
In this case, a link is  obtained between
the proton width and the squared ANC  of the mirror neutron state.
The relation between mirror overlaps  can be used
to study astrophysically relevant proton  capture reactions based on information 
obtained from transfer reactions with stable beams.   

\end{abstract}

\maketitle

The astrophysical $S$-factor associated with the peripheral
proton capture reaction 
$B(p,\gamma)A$ at stellar energies is well known\cite{xu} to be related to the 
Asymptotic Normalization Coefficient (ANC) of the virtual decay 
$A \rightarrow B + p$. The same ANCs play a crucial role in other 
peripheral processes such as  transfer 
reactions whose cross sections
are significantly higher and therefore more easily measurable than those of the 
direct capture processes at
astrophysically relevant energies \cite{xu}. The  study
of ANCs of astrophysical interest is a new and rapidly developing
direction in modern experimental nuclear physics \cite{Cas00,Kub02}. 
 However, in order to exploit these ideas to determine the ANCs for light 
proton-rich 
nuclei 
 of importance to nuclear astrophysics the corresponding transfer reactions  
often require the use of
weak radioactive beams  which generally involves more difficult and less 
accurate experiments than are possible with stable beams.
 The higher intensities of stable beams means
that they can be used at  
energies below the Coulomb barrier where the sensitivity 
to   optical potentials, which are the main uncertainty of
ANCs determined from  transfer reactions, is minimised. We point out here that 
the ANC of the virtual neutron decay of the nucleus mirror to $A$, which may be 
susceptible to study with stable beams, is related in a model independent way by 
the charge symmetry of nuclear forces to the ANC of the corresponding proton 
decay of $A$. We propose to exploit this new insight to predict  peripheral
reaction cross sections in stars.

An asymptotic normalization coefficient (ANC) is one 
of the fundamental characteristics of the virtual
decay of a nucleus into two clusters  
and is equivalent
to the coupling constants in particle physics \cite{BBD77}. 
When multiplied by  a trivial factor, it equals to the on-shell
amplitude for the virtual decay into two clusters
 and it determines the large 
distance behaviour of the projection of the  bound state
wave function of the nucleus onto a binary %two-body 
channel.

In earlier %theoretical 
work \cite{Tim,TI}, the ANCs for the  one-nucleon 
virtual decays of the
mirror pairs $^8$B $-$ $^8$Li
and $^{12}$N $-$ $^{12}$B were studied 
in a microscopic approach. The calculated ANCs themselves
 depended strongly on the choice of the nucleon-nucleon (NN) force but the 
ratios of ANCs for mirror pairs
were  practically
independent of the choice of the NN force. 
This observation is based so far entirely on the calculations using detailed 
models of nuclear structure. We now show that it follows naturally as a 
consequence of the charge symmetry of nuclear forces \footnote{We believe that 
charge symmetry rather than full charge independence is involved because mirror 
nuclei have the same number of n-p pairs.}.  

\begin{table*}
\caption{Ratio ${\cal R}$ of the proton and neutron squared ANC's  %$C_{lj}$ 
for mirror overlap integrals calculated with  GPT \cite{GPT},
Volkov  V1 \cite{Volk}, Brink-Boeker  B1 \cite{BB}
and four versions of the M3Y effective NN potentials \cite{M3Y,Par}. 
M3Y(R), M3Y(HJ) and M3Y(P) were fitted to the oscillator G-matrix
elements of the Reid, Hamada-Johnston and  Paris NN potentials  respectively
and M3Y(E) was fitted to
the oscillator G-matrix elements derived from the NN scattering data.
Analytical estimates ${\cal R}_0$ are also shown.} 
\begin {center}
\begin{tabular}{ ccc c c c c c ccc}
\hline  
Overlap & Mirror overlap & $j$ &GPT &V1&B1&M3Y(E)&M3Y(R)&M3Y(P)&M3Y(HJ)&
${\cal R}_0$ \\
\hline
$\la^{6}Li|^{7}Be\ra$ & $\la^{6}Li|^{7}Li\ra$ &
  1/2 & 1.05 & 1.05 & 1.05 & 1.05 & 1.05 & 1.05 & 1.05& 1.05\\
 & & 3/2 & 1.05 & 1.05 & 1.05 & 1.05 & 1.05 & 1.05 & 1.05& 1.05\\
$\la^{7}Be|^{8}B\ra$ & $\la^{7}Li|^{8}Li\ra$ &
  1/2 & 1.18 & 1.17 & 1.18 & 1.20 & 1.17 & 1.17 & 1.19& 1.12\\
 & & 3/2 & 1.20 & 1.20 & 1.20 & 1.22 & 1.19 & 1.20 & 1.22& 1.12\\
$\la^{11}C|^{12}N\ra$ & $\la^{11}B|^{12}B\ra$ &
  1/2 & 1.45 & 1.42 & 1.44 & 1.45 & 1.42 & 1.42 & 1.43 & 1.37\\
 & & 3/2 & 1.45 & 1.44 & 1.46 & 1.46 & 1.42 & 1.43 & 1.46& 1.37\\
$\la^{12}C|^{13}N\ra$ & $\la^{12}C|^{13}C\ra$ &
  1/2 & 1.26 & 1.24 & 1.24 & 1.26 & 1.23 & 1.24 & 1.25& 1.19\\
$\la^{14}N|^{15}O\ra$ & $\la^{14}N|^{15}N\ra$ &
  1/2 & 1.53 & 1.50 & 1.51 & 1.53 & 1.51 & 1.51 & 1.52& 1.48\\
 & & 3/2 & 1.54 & 1.51 & 1.51 & 1.57 & 1.51 & 1.52 & 1.56& 1.48\\
$\la$ $ ^{15}N|^{16}O\ra$ & $\la^{15}O|^{16}O\ra$ &
  1/2 & 1.55 & 1.54 & 1.54 & 1.57 & 1.54 & 1.55 & 1.56& 1.52\\ 
\hline
\end{tabular}
\end{center}
\label{table1}
\end{table*} 

%\end{document}

The ANC $C_{lj}$ for the one-nucleon virtual decay $A \rightarrow B + N$ 
is defined  via the tail of  the overlap integral $I_{lj}(r)$
between the wave functions of nuclei $A$ and $B = A - 1$, where
$l$ is the  orbital momentum and $j$ is the total relative angular momentum
between $B$ and $N$. 
Asymptotically, this  overlap behaves as
\beq
\sqrt{A}\,I_{lj}(r)\approx C_{lj}\frac{W_{-\eta,l+1/2}(2\kappa r)}
{r}, 
\,\,\,\,\,\, 
r\rightarrow\infty,
\eeqn{anc}
where $\kappa =(2\mu\epsilon/\hbar^2)^{1/2}$, $\epsilon$ is the one-nucleon
separation energy, $\eta=Z_BZ_Ne^2\mu/\hbar^2\kappa $, 
$\mu$ is the reduced mass for the $B + N$ system and $W$ is the Whittaker 
function. It follows from %Refs. 
\cite{BBD77,MT90,Tim} that  $C_{lj}$ 
can be expressed in terms of the many-body wave functions of the nuclei $A$ 
and $B$: 
%through the expression
\beq
 C_{lj}  &=& -\frac{2\mu \sqrt{A}}{\hbar^2}\nonumber \\
 &\times & 
\la[[ \varphi_l(i\kappa r)
Y_{l}(\hat{\ve{r}}) \otimes \chi_{\frac{1}{2}} ]_j
\otimes \Psi_{J_B}]_{J_A}||\hat{{\cal V}}||\Psi_{J_A}\ra, 
\eeqn{vff}
where 
\beq
\varphi_l(i\kappa r) = 
 e^{-i\sigma_l} F_l(i\kappa r)/{\kappa r}, 
\eeqn{phi}
$F_l$ is the regular Coulomb wave function at imaginary momentum
$i\kappa$, $\sigma_l = {\rm arg} \,\Gamma(l+1+i\eta)$, $r$ is the distance 
between $N$  and the center-of-mass of $B$  and
\beq
\hat{{\cal V}} = 
\sum_{i=1}^{A-1} V_{NN} (|\ve{r}_i-\ve{r}_A|)+\Delta V_{coul} =
\hat{{\cal V}}_N+\Delta V_{coul},
\eeqn{V}
\beq
\Delta V_{coul} = \sum_{i=1}^{A-1} 
\frac{e_ie_A} {|\ve{r}_i-\ve{r}_A|} -\frac{Z_Be_Ae}{r}. 
\eeqn{dV}
Here $e_i$ ($e_A$) is the charge of the $i$-th ($A$-th)
nucleon, $Z_B$ is the charge of the residual nucleus $B$ 
and $V_{NN}$ is the two-body nuclear NN potential.
If the separated nucleon is  a neutron,
$\varphi_l$ is replaced by the Bessel function $j_l(i\kappa r)$.

ANCs can be obtained  from Eq. (\ref{vff}) using  
wave functions  which model the structure of
nuclear interior well, for example,  from the oscillator shell model
\footnote{%Assumption about oscillator radial shape 
The oscillator shape of the single-particle
wave functions makes the correct
treatment of the center-of-mass possible, which is crucial for the
nuclei considered here.}.
The incorrect behavior of these model wave functions at large distances plays a 
minor role because of the presence of the 
short range NN potential on the right hand side in Eq. (\ref{vff}) \cite{MT90}.
We have performed such
calculations for several  0$p$ nuclei, some of which are of astrophysical
importance,  with a range of NN potentials using fixed
0$\hbar\omega$  wave functions obtained in \cite{Boyar},\footnote{
%Most shell model calculations deal with sets of numbers representing matrix 
%elements of the NN interaction.
%% and do not asume any radial dependence of the 
%%single-particle wave functions. 
%Eq. (2) requires 
%components of the NN interactions \cite{MT90}, which are not present
%in the Hamiltonians used to obtain the wave functions of nuclei $A$ and $B$.
%Here we use fixed known
%shell model wave functions 
%%(we have chosen the wave functions, tabulated in
%%\cite{Boyar}, which give good positions of the energy levels of the 
%%0$p$-nuclei)
%and  calculate the two-body matrix elements entering  
%%for the amplitude of the virtual decay in 
%Eq.(2) by using 
%NN potentials from the literature.
The more consistent approach used in
\cite{Tim}, in which the same
NN potential is used both in Eq. (2) and in the shell-model Hamiltonians,
leads to similar results for the ratio of the mirror ANCs.}. 
The oscillator radii was chosen to 
provide correct sizes for the nuclei considered. 
In these calculations
  mirror nuclei have exactly the same wave functions but, of course, 
%Eq. (\ref{vff}) predicts different ANCs because of the presence of the Coulomb 
%function in the case of virtual proton decay.
the mirror ANCs are different because of different functions $\varphi_l(i\kappa 
r)$
  involved.
%The  calculated ANC's   strongly depend on the  NN potential choice.
The  $|C_{lj}|^2$ values change by a factor of two for different
NN potential choice, but
the ratio ${\cal R} =|C_p/C_n|^2$, where $C_p$ and $C_n$ are the proton 
and neutron 
ANCs for mirrors and hence may refer to different nuclei, 
changes  by less than 4$\%$ for each 
mirror pair of  overlaps (see Table 1 and
\footnote[1]{We have found a mistake in the computer code
for proton ANC's used in Refs. \cite{Tim,TI}. The corrected values
are given in Table 1.}).

The observed effect has the following explanation.
We first replace $\Delta V_{coul}$ by $V_{coul}(r)-Z_Be_A e/r$
where $V_{coul}(r)$ is the monopole Coulomb interaction of the $A$-th
nucleon with the nucleus $B$.
This ignores higher multipole components of $\Delta V_{coul}$. Eq. (\ref{vff})
can then be replaced exactly by a formula in which $\Delta V_{coul}$
is removed from the matrix element and $\varphi _l(r)$ is replaced by
$\varphi _l^{mod}(r)$. The latter is defined as  the regular solution of the 
Schr\"odinger equation with the potential $V_{coul}(r)$ and which is normalized 
so that 
$\varphi_l(r) = \varphi _l^{mod}(r)$ outside the charge radius of $B$. 
Inside the charge radius, the potential $V_{coul}(r)$ varies little over 
the nuclear volume and can be replaced by a constant equal to the 
separation  energies difference
$\epsilon_n - \epsilon_p$. Hence, in the nuclear interior $r < R_N$,
which is all that matters on the right-hand-side of Eq. 
(\ref{vff}), we can use
\begin{equation}
\varphi _l^{mod}(r)=
\frac{F_l(i\kappa _{p} r_{N})}{\kappa _{p} R_{N}j_l(i\kappa _{n} 
R_{N})}j_l(i\kappa _{n} r),\,\,\,\,\,r\leq R_{N}, \label{phi1}\end{equation}
where $i\kappa _{p}$ and $i\kappa _{n}$ are determined by the proton 
and neutron 
separation energies $\epsilon_p$ and $\epsilon_n$. 
Using Eq. (\ref{phi1}) in the modified Eq. (\ref{vff}) and making 
the assumption 
that the difference between the wave functions for mirror pairs in the nuclear 
interior  can be ignored,  we find
\beq
{\cal R } \approx {\cal R}_0 = \left|\frac{F_l(i\kappa_pR_N)}
{\kappa_pR_N\,j_l(i\kappa_nR_N)}\right|^2.
\eeqn{rn}
${\cal R}_0 $ depends on the $NN$ force only implicitly through $R_{N}$.

The values of ${\cal R}_0$, presented in Table I, have been calculated for 
$R_N = 1.3\cdot B^{1/3}$. They change by less than 2$\%$, 
when $R_N$ is varied  from 2.5 to 4.5 fm in each case, and
are smaller by less than $7\%$
than the ${\cal R}$ values obtained from microscopic calculations. 
 Eq. (\ref{rn}) correctly predicts the dependence of $\cal{R}$ on neutron and 
proton
separation energies. The tendency of ${\cal R}_0$ to inderestimate 
${\cal R}$ can be attributed to the contributions from the
$r^{-2}$ and $r^{-3}$ multipoles of $\Delta V_{coul}$. When these multipoles
are excluded from the microscopic calculations, the ${\cal R}$ values
decrease and become equal to ${\cal R}_0$ within the uncertainty in its
definition.

\begin{table*}
\caption{Squared ratio   $(b_{max}/b_{min})^2$
of the maximal and minimal values of $b$,
average ratio of squared ANC's  $\la {\cal R}_b\ra$
analytical estimates ${\cal R}_0$ and
  experimental ratios ${\cal R}^{exp}$. 
Where several experimental values of ANC's are available, 
we take their average. Also shown are  proton  $\epsilon_p$ and neutron 
$\epsilon_n$ separation energies (in MeV), number of nodes $n$ and
orbital momentum $l$.
 } 
\begin {center}
\begin{tabular}{ ccccc c c c  ccc}
\hline 
\\ 
Overlap & $\epsilon_p$ & Mirror overlap &  $\epsilon_n$  &  $\,nl\,$ 
& $\left(\frac{b_{max}}{b_{min}}\right)^2$ & $\la {\cal R}_b\ra$ &
 ${\cal R}_0$ & ${\cal R}^{exp}$ & 
 $\begin{array}{c} {\rm Ref. for} \\ C_p^{exp} \end{array}$ & 
 $\begin{array}{c} {\rm Ref. for } \\ C_n^{exp} \end{array}$\\
\\
\hline
$\la^{7}Be|^{8}B\ra$ & 0.137 & $\la^{7}Li|^{8}Li\ra$ & 2.033 &
  $0p$ &   1.23 & 1.01$\pm$0.01    & 1.12& 1.08 $\pm$ 0.15
  & \cite{azh} & \cite{livius}\\
$\la^{11}C|^{12}N\ra$ & 0.601 & $\la^{11}B|^{12}B\ra$ &3.370 &
  $0p$ &   1.67 &1.30$\pm$0.02 &   1.37 &1.28 $\pm$ 0.29 & 
  \cite{XT} & \cite{Liu}\\
$\la$ $^{14}N|^{15}O(\frac{3}{2}_1^+)\ra$ & 0.507 &
$\la^{14}N|^{15}N(\frac{3}{2}_1^+)\ra$ & 3.026 &
  $1s$ &  1.68 & 3.62$\pm$0.03 & 4.09   && &\\ 
$\la$ $^{15}N|^{16}O\ra$ & 12.128 & $\la^{15}O|^{16}O\ra$ & 15.664 &
  $0p$ &   2.55 & 1.55$\pm$0.02 &1.52 &  & &\\ 
$\la^{16}O|^{17}F(\frac{5}{2}_1^+)\ra$ & 0.601 &
$\la^{16}O|^{17}O(\frac{5}{2}_1^+)\ra$ & 4.144 &
    $0d$ &   2.15 & 1.21$\pm$0.03 & 1.21 &1.33 $\pm$ 0.20
  & \cite{fort,vern,art,gagl} & \cite{bz}\\ 
$\la^{16}O|^{17}F(\frac{1}{2}_1^+)\ra$ & 0.106 &
$\la^{16}O|^{17}O(\frac{1}{2}_1^+)\ra$ & 3.273 &
  $1s$ &   1.56 & 702$\pm$4 &   796 &   & &\\ 
 $\la^{22}Mg|^{23}Al\ra$ & 0.123 & $\la^{22}Ne|^{23}Ne\ra$ & 4.419 &
  $0d$ &   1.50 & 2.67$\cdot 10^4$  &  2.61$\cdot 10^4$ &   &   &  \\ 
$\la^{26}Si|^{27}P\ra$ & 0.859 & $\la^{26}Mg|^{27}Mg\ra$ & 6.443 &
  $1s$ &   1.80 & 40.3$\pm$1.1 &  43.3  &   & &\\ 
  \hline
\end{tabular}
\end{center}
\label{table2}
\end{table*}

In practice, overlap integrals for transfer reactions are frequently modelled
as normalised single-particle wave functions times  spectroscopic factors $S$,
so that $C_{p(n)} = \sqrt{S_{p(n)}}\, b_{p(n)}$, where $b_{p(n)}$ 
is the single-particle proton (neutron)  ANC. 
%This approach leads to an  
%alternative way to estimate ${\cal R}$
The derivation above shows that the result Eq.( \ref{rn}) is valid for $\mid 
b_p/b_n \mid ^2$ if we  assume that the single particle wave functions in the
interior and the 
nuclear single particle potentials are the same for p and n. The ratio 
${\cal R}_b = (b_p/b_n)^2$ is 
therefore expected to have only weak dependence on these potentials. 
We have verified this for 
a range of potentials chosen to simultaneouly
reproduce fixed proton and neutron binding 
energies. The individual ANC's $b_n$ and $b_p$ vary by up to a factor of 2 but 
the ratio ${\cal R}_b$ is stable to within 3$\%$ with an 
average which agrees  with Eq.(\ref{rn})\footnote{The difference between 
the average $<{\cal R}_b>$ and ${\cal R}_0$ is largest for weakly bound states. 
We will return to the reason for this elsewhere.}. If we %make the additional 
%assumption 
assume that the spectroscopic factors $S_p$ and $S_n$ are equal for mirror 
pairs then we have an alternative way of estimating $\cal R$. Note however that 
our derivation of Eq.(\ref{rn}) involves fewer assumptions than in this 
alternative approach and in fact does not appeal to the concept of spectroscopic 
factor at all. Our approach is therefore much more general and provides a basis 
for further refinement of the value of the ratio ${\cal R}$ predicted by theory. 
For the mirror pairs $^8$B - $^8$Li, $^{12}$N - $^{12}$B and 
$^{17}$F - $^{17}$O, where the experimental values of the
proton $C_p^{exp}$ and neutron $C_n^{exp}$ ANC's are simultaneously available, 
both $\la {\cal R}_b\ra $ and ${\cal R}_0$ agree with
${\cal R}^{exp} = |C_p^{exp}/C_n^{exp}|^2$  within the error bars 
(see Table II).

Near the edge of stability, where  neutron separation energies 
become very small, the corresponding mirror proton states  
%have positive energies and are observed 
manifest themselves as resonances. 
The width $\Gamma_p$ of a narrow proton resonance is related to the  
ANC  of the Gamow wave function for this resonance by the equation
$\Gamma_p = \mu/\kappa_p |C_p|^2$
\cite{MTr}.
The ANC $C_p$ can be calculated from Eqs. 
(\ref{vff}) and (\ref{phi}) using  the regular Coulomb function
$F_l(\kappa_p r)$ of a real argument \cite{BBD77}.
 Therefore,  a link must exist between $\Gamma_p$ %the proton width 
 and the ANC of mirror neutron bound states. The ratio 
${\cal R}_{\Gamma} = \Gamma_p/|C_n|^2$ is then approximated by an equation
 similar to Eq. (\ref{rn}):
\beq
{\cal R}_{\Gamma} \approx {\cal R}_0^{res} = 
\frac{\kappa_p}{\mu}
\left|\frac{F_l(\kappa_pR_N)}{\kappa_pR_N\,j_l(i\kappa_nR_N)}\right|^2
\eeqn{rnres}

Alternatively, ${\cal R}_{\Gamma}$ can be approximated by the single-particle
ratio ${\cal R}_{\Gamma}^{s.p.} = \Gamma_p^{s.p.}/b_n^2$ 
if the spectroscopic factors
and  single-particle potential wells for mirror bound-unbound
pairs are assumed equal. We have
calculated %the ratios 
${\cal R}_{\Gamma}^{s.p.}$ 
 for the $^8$B(1$^+$), $^{12}$N(2$^+$),
$^{13}$N($\frac{1}{2}^+$) and $^{13}$N($\frac{5}{2}^+$) resonances
using a set of two-body Woods-Saxon potentials which reproduce
both the separation energy of the loosely-bound neutron and the position of 
the mirror proton resonance.
In the case of $l \neq 0$, for different choice of the two-body potentials
 the ratios ${\cal R}_{\Gamma}^{s.p.}$ change by about 3$\%$
while   $\Gamma_p^{s.p.}$ changes by up to a factor of 2 (see Table III).
This is the same as in the case of bound mirror pairs of overlaps.
However, for $l = 0$, where the centrifugal barrier is absent and
non-resonant contributions are larger, the change
in ${\cal R}_{\Gamma}^{s.p.}$ is  larger and reaches 11$\%$.
The average value of ${\cal R}_{\Gamma}^{s.p.}$ agrees with ${\cal R}_0^{res}$
for the $l = 2$ resonance $^{13}$N($\frac{5}{2}^+$)  but
is smaller than ${\cal R}_0^{res}$ by 16$\%$, 20$\%$ and 37$\%$ for
$^8$B(1$^+$), $^{12}$N(2$^+$) and $^{13}$N($\frac{1}{2}^+$) respectively.
The ${\cal R}_0^{res}$ values themselves 
are quite stable with respect to different choice of
$R_N$ except in the case of the $l = 0$ resonance
$^{13}$N($\frac{1}{2}^+)$ where the uncertainty
of ${\cal R}_0^{res}$ is $5\%$.

\begin{table*}
\caption{The ratio   $\Gamma_p^{max}/\Gamma_p^{min}$ of
the maximal and minimal proton widths,
average ratio of ${\cal R}_{\Gamma}^{s.p.}$, 
analytical estimates ${\cal R}_0^{res}$ and
 experimental ratios  ${\cal R}^{exp}_{\Gamma}$.
Where several experimental values of ANC's are available, 
we take their average.} 
\begin {center}
\begin{tabular}{ cc c c cc ccc}
\hline 
\\ 
Proton resonance & $\,\,\,$
$\begin{array} {c} {\rm Bound \, mirror} \\ {\rm analog} \end{array} $
$\,\,\,$ & $\,\,\,\,\,\,l\,\,\,\,\,\,$ & 
 $\,\,\,\frac{\Gamma^{max}_p}{\Gamma^{min}_p}\,\,\,$ & 
$\,\,\la {\cal R}_{\Gamma}^{s.p.}\ra\,\,$ & ${\cal R}_0^{res}$ &
${\cal R}_{\Gamma}^{exp}$ & $\begin{array} {c} 
{\rm Ref. for} \\ \Gamma_p^{exp} \end{array} $ &
$\begin{array} {c} 
 {\rm Ref. for} \\ C_p^{exp} \end{array}$\\
\\
\hline
$^8$B(1$^+$, 0.774) &  $^8$Li(1$^+$, 0.980) & 1 & 
    1.43 &$(1.70\pm0.03)\cdot 10^{-3}$ &  2.03$\cdot 10^{-3}$ &
   (2.29 $\pm$ 0.40)$\cdot 10^{-3}$ & \cite{Aiz8} & \cite{livius}\\ 
$^{12}$N(2$^+$, 0.960) &  $^{12}$B(2$^+$, 0.953) & 1 &
   1.61 &$(1.22\pm 0.01)\cdot 10^{-5}$ & 1.42$\cdot 10^{-5}$ && &\\ 
$^{13}$N($\frac{1}{2}^+$, 2.36) &  $^{13}$C($\frac{1}{2}^+$, 3.09) & 0 & 
  1.55 &   $(5.98\pm 0.32)\cdot 10^{-5}$&  $8.5\cdot 10^{-5}$&
 (4.57 $\pm$ 0.57)$\cdot 10^{-5}$ 
 & \cite{Aiz13,Ima01} & \cite{Liu}\\ 
 $^{13}$N($\frac{5}{2}^+$, 3.55) &  $^{13}$C($\frac{5}{2}^+$, 3.85) & 2 & 
   2.01&  $(1.37\pm 0.03)\cdot 10^{-2}$ &
  1.42$\cdot 10^{-2}$ & (1.06 $\pm$ 0.21)$\cdot 10^{-2}$
& \cite{Aiz13} & \cite{Liu}  \\ 
 \hline
 \end{tabular}
\end{center}
\label{table3}
\end{table*}

The  proton widths of $^8$B(1$^+$)
$^{13}$N($\frac{1}{2}^+$)
and $^{13}$N($\frac{5}{2}^+$)
and neutron ANC's for their
mirror states are known experimentally. The ratios
${\cal R}_{\Gamma}^{exp} = \Gamma^{exp}_p/|C_n^{exp}|^2$ 
for these states are shown in Table III. 
In all these cases, the %experimental data do not confirm a 
single-particle approximation
${\cal R}_{\Gamma} \approx {\cal R}_{\Gamma}^{s.p.}$
is not confirmed. For $^8$B(1$^+$),
${\cal R}_{\Gamma}^{exp}$ is larger than ${\cal R}_{\Gamma}^{s.p.}$
and agrees with ${\cal R}_0^{res}$, but for $^{13}$N($\frac{1}{2}^+$)
and $^{13}$N($\frac{5}{2}^+$) %the ratios 
${\cal R}_{\Gamma}^{exp}$ is
significantly lower than  ${\cal R}_{\Gamma}^{s.p.}$ and ${\cal R}_0^{res}$.
This result suggests that estimates based on the relation
$\Gamma_p = S_p \Gamma_p^{s.p.}$
and the assumption $S_p = S_n$ can be unreliable.

The present work confirms the existence of a link between proton and neutron 
mirror ANC's both for bound-bound and bound-unbound mirror pairs.  
Therefore,  neutron ANC's obtained with stable beams
can be used to predict cross sections of  low-energy  direct and resonance
proton capture reactions. 
Although  more accurate theoretical ratios for ${\cal R}$ and 
${\cal R}_{\Gamma}$ are required for these purposes, 
the estimates $\la {\cal R}_b \ra$, 
${\cal R}_0$, ${\cal R}_{\Gamma}^{s.p.}$ and ${\cal R}_0^{res}$ 
of the present paper 
can already be used in some cases.
In fact, the ratio ${\cal R}$ has already been  used to predict 
the direct $^{11}$C($p,\gamma)^{12}$N capture cross sections
in  \cite{TI} and the results obtained there are 
in a good agreement with the predictions based on proton ANC's recently
measured in \cite{XT}. Also, the astrophysical S-factor for the
$^7$Be($p,\gamma)^8$B reaction has been calculated 
in  \cite{livius} based on the $\la {\cal R}_b \ra$ estimate and
experimentally measured neutron ANC in $^8$Li. 
Another example is the proton width of the $^{12}$N$(2^+)$ resonance
for which
only an upper limit of 20 keV is available. Using the neutron
ANC for the mirror $^{12}$B$(2^+)$ state from  \cite{Liu}, we can predict
that $\Gamma_p$ is equal to $5.9 \pm 1.0$ or $6.9 \pm 1.2$ keV for the
${\cal R}_{\Gamma} \approx
\la {\cal R}_{\Gamma}^{s.p.}\ra$ and ${\cal R}_{\Gamma} \approx  
{\cal R}_0^{res}$ assumptions
respectively.  These values are less uncertain
than the currently available experimental limit $\Gamma_p < 20$ keV.

Among other cases of astrophysical interest is 
the astrophysical $S$-factor for the direct capture reaction
$^{14}{\rm N}(p, \gamma)^{15}{\rm O}(\frac{3}{2}_1^+)$, 
 which is mainly responsible for the energy production in the CNO cycle.
The $^{15}O(\frac{3}{2}_1^+)$ state is separated from the
neighbouring   $^{15}O(\frac{5}{2}_2^+)$ state  by only 70 KeV, which 
influences the precision of  measurements involving this state.
The spacing between the mirror $^{15}N(\frac{5}{2}_2^+)$
and $^{15}N(\frac{3}{2}_1^+)$ states is larger and therefore
the ANC for the
$\la {}^{14}{\rm N}|{}^{15}{\rm N}(\frac{3}{2}_1^+)\ra$
overlap integral can be determined using neutron transfer reactions
to higher accuracy than the 
$^{15}O(\frac{3}{2}_1^+)$ ANC.
%The $\la {}^{14}{\rm N}|{}^{15}{\rm N}(\frac{3}{2}_1^+)\ra$ ANC can be 
%.
 Also, direct contributions to the cross sections of  the
$^{22}$Mg($p,\gamma)^{23}$Al and
$^{26}$Si($p,\gamma)^{27}$P reactions, involving proton-rich radioactive
nuclei, could be calculated through the mirror neutron ANC's which can
be determined using stable targets $^{22}$Ne  and $^{26}$Mg.
These reaction  are relevant to the nucleosynthesis in novae and
are being intensively investigated.

This work has been supported by the UK EPSRC grant GR/M/82141,
 the U.\,S. Department of Energy
under Grant No.\@ DE-FG03-93ER40773 and the U.\,S. National Science
Foundation under Grant No.\ PHY-0140343.
 N.T. is grateful to B.V. Danilin for some useful discussions.

\end{document}